\documentclass{jpp}

\usepackage[utf8]{inputenc}
\usepackage[T1]{fontenc}
\usepackage{amsfonts,amsmath,amssymb}
\usepackage{bm,nicefrac}
\usepackage{graphicx}
\usepackage{tabularx}
\usepackage{subcaption}
\usepackage[section]{placeins}
\usepackage{textgreek}
\usepackage[breaklinks=true,colorlinks=true,linkcolor=blue,urlcolor=blue,citecolor=blue]{hyperref} 
\usepackage[table]{xcolor}
\usepackage[english]{babel}
\usepackage{listings}
\usepackage{algorithm}
\usepackage{algpseudocode}

\DeclareMathOperator*{\argmin}{arg\,min}

\title{Stellarator Optimization with Constraints}
\author{
Rory Conlin \aff{1}  \corresp{\email{wconlin@princeton.edu}}, 
Patrick Kim \aff{2},
Daniel W. Dudt \aff{1}, 
Dario Panici \aff{1},
\and Egemen Kolemen \aff{1} \corresp{\email{ekolemen@princeton.edu}}}

\affiliation{\aff{1}Princeton University, Princeton, New Jersey 08544}

\begin{document}

\maketitle

\begin{abstract}
    In this work we consider the problem of optimizing a stellarator subject to hard constraints on the design variables and physics properties of the equilibrium. We survey current numerical methods for handling these constraints, and summarize a number of methods from the wider optimization community that have not been used extensively for stellarator optimization thus far. We demonstrate the utility of new methods of constrained optimization by optimizing a QA stellarator for favorable physics properties while preventing strong shaping of the plasma boundary which can be difficult to create with external current sources.
\end{abstract}

\section{Introduction}
Stellarator optimization is naturally full of constraints that must be satisfied by an optimized configuration. Many of these constraints are due to engineering requirements, such as reasonable aspect ratio, minimum coil-plasma distance, or maximum curvature of the coils or plasma boundary. Others come from physics requirements, such as having a prescribed rotational transform at the boundary for an island divertor \citep{feng2006islanddivertor}, having a stable magnetic well \citep{landreman2020magneticwell}, or ensuring the net current is zero. There are also constraints to enforce self-consistency between different models of the plasma, such as requiring that the prescribed current in an ideal MHD equilibrium solver matches with the actual bootstrap current from a kinetic calculation \citep{landreman2022bootstrap}, or that the coils correctly cancel the normal field on the plasma boundary. Perhaps the most important constraint is that any optimized configuration must actually be in MHD equilibrium.

A generic optimization problem can be written as:

\begin{equation}
\begin{aligned}
    \min_\mathbf{x} \quad & f(\mathbf{x}) \\
\textrm{s.t.} \quad &     \mathbf{g}_E(\mathbf{x}) = 0 \\
  &\mathbf{l} \le \mathbf{g}_I(\mathbf{x}) \le \mathbf{u}    \\
\end{aligned}
\end{equation}

where $\mathbf{x} \in \mathcal{R}^n$ is a vector of decision variables, $f : \mathcal{R}^n \rightarrow \mathcal{R}^1$ is the objective function, $\mathbf{g}_E : \mathcal{R}^n \rightarrow \mathcal{R}^{m_1}$ are equality constraints, $\mathbf{g}_I : \mathcal{R}^n \rightarrow \mathcal{R}^{m_2}$ are inequality constraints, and $\mathbf{l},\mathbf{u} \in \mathcal{R}^{m_2}$ are lower and upper bounds on the inequality constraints (which may be infinite for one sided constraints). By introducing slack variables $\mathbf{s}$, we can transform the inequality constraint into an equality constraint and simple bounds:

\begin{equation}
\begin{aligned}
    \min_\mathbf{x,s} \quad & f(\mathbf{x}) \\
\textrm{s.t.} \quad &     \mathbf{g}_E(\mathbf{x}) = 0 \\
   & \mathbf{g}_I(\mathbf{x}) - \mathbf{s} = 0    \\
   & \mathbf{l} \le \mathbf{s} \le \mathbf{u} \\
\end{aligned}
\end{equation}

Which if we absorb $\mathbf{s}$ into $\mathbf{x}$ and combine the constraints into a single function $\mathbf{g}$, we can write in standard form as:

\begin{equation}
\begin{aligned}
    \min_\mathbf{x} \quad & f(\mathbf{x}) \\
\textrm{s.t.} \quad &     \mathbf{g}(\mathbf{x}) = 0 \\
   & \mathbf{l} \le \mathbf{x} \le \mathbf{u} \\
\end{aligned}
\label{eq:opt1}
\end{equation}

In \autoref{sec:existing} we will summarize methods for handling constraints that are already in use in stellarator optimization such as linear constraint projection and sum of squares. In \autoref{sec:constrained} we introduce several methods from the wider optimization literature and discuss their properties and relevance for the stellarator optimization field. In \autoref{sec:indesc} we detail the implementation of a new augmented Lagrangian optimization algorithm into the DESC code, and demonstrate it's use with an example problem in \autoref{sec:examples}.

\section{Existing methods}
\label{sec:existing}
\subsection{Linear Equality Constraints}

Constraints where the decision variables $\mathbf{x}$ appear only linearly can be handled efficiently by suitably redefining variables. A problem with linear constraints can be written as:

\begin{equation}
\begin{aligned}
    \min_{\mathbf{x} \in R^n} \quad & f(\mathbf{x}) \\
\textrm{s.t.} \quad &     A \mathbf{x} - \mathbf{b} = 0 \\
\end{aligned}
\end{equation}

Where $A \in R^{m \times n}$ and $\mathbf{b} \in R^m$, $m \le n$ are the coefficients that define the linear constraint. We can then decompose $\mathbf{x} = \mathbf{x}_p + Z \mathbf{y}$ where $\mathbf{x}_p \in R^n$ is any particular solution to the under-determined system $A \mathbf{x} = \mathbf{b}$ (typically taken to be the least-norm solution), $Z \in R^{n \times (n-m)}$ is a (typically orthogonal) representation for the null space of $A$, such that $AZ=0$, and $\mathbf{y} \in R^{n-m}$ is an arbitrary vector. This allows us to write the problem as:

\begin{equation}
\begin{aligned}
    \min_{\mathbf{y} \in R^{n-m}} \quad & f(\mathbf{x}_p + Z\mathbf{y}) \\
\end{aligned}
\end{equation}

which is now an unconstrained problem, where any value for $\mathbf{y}$ is feasible, and the full solution $\mathbf{x}^*$ can be recovered as $\mathbf{x}_p + Z \mathbf{y}^*$.

While simple to deal with, linear equality constraints are very limited in their applicability. Most constraints that are of interest in stellarator optimization are nonlinear, making linear equality constraints only useful for enforcing simple boundary conditions on the plasma boundary and profiles.

\subsection{Sum of Squares}

Perhaps the simplest way to (approximately) solve \autoref{eq:opt1} is with a quadratic penalty method, where the objective and constraints are combined into a single function, which is then minimized without constraints:

\begin{equation}
    \min_\mathbf{x} \quad f(\mathbf{x}) + w_1 ||\mathbf{g}(\mathbf{x}) ||^2 + w_2 || I_{(l,u)}(\mathbf{x}) ||^2
    \label{eq:quad_penalty}
\end{equation}

Where $w_1$ and $w_2$ are weights, and $I_{(l,u)}(\mathbf{x})$ is a "deadzone" function:
\begin{equation}
I_{(l,u)}(\mathbf{x}) = 
\left\{
    \begin{array}{lr}
        0 & \text{if } l \le \mathbf{x} \le u \\
        \max (u-\mathbf{x}, \mathbf{x}-l)& \text{otherwise } 
    \end{array}
\right\}
\end{equation}

This formulation is used by a number of codes for optimizing stellarator equilibria, such as SIMSOPT, STELLOPT, ROSE, and DESC. While simple to implement, this formulation has a number of drawbacks:

\begin{enumerate}
    \item The weights $w$ must be determined by the user, often through extensive trial and error 
    \item The constraints are only satisfied as $w \rightarrow \infty$, leading to a badly scaled problem. 
    \item The $\max$ term in the deadzone function introduces a discontinuity in the second derivative of the objective, which can hinder progress of some optimization methods that assume the objective and constraints to be at least C2.
\end{enumerate}

\subsection{Projection Methods}

The constraint of MHD equilibrium is most commonly dealt with by using a fixed boundary equilibrium solver at each step in the optimization. Given the plasma boundary and profiles at the latest optimization step, the equilibrium solver is run to determine the full solution throughout the plasma volume, and this is then used to evaluate the objective function. This is akin to a projection method, where after each step of the optimizer, the solution is projected back onto the feasible region $\{\mathbf{x} ~ | ~ \mathbf{g}(\mathbf{x})=0\}$. This also suffers from a number of drawbacks:
\begin{enumerate}
    \item The equilibrium must be re-solved at each step of the optimizer, possibly multiple times if finite differences are used to evaluate derivatives. This can be made somewhat more efficient if a warm start is used based on the previous accepted step \citep{conlin2023desc, dudt2023desc}, though this is still often the most expensive part of the optimization loop. 
    \item Projection type methods are limited in applicability to constraints for which a projection operator can be defined. MHD equilibrium solvers can serve this function for equilibrium constraints, but a projection operator for a general nonlinear constraint is not known.
    \item Enforcing that the constraints are satisfied at each step of the optimization may be too strict, and cause the optimizer to stop at a poor local minimum. 
\end{enumerate}

\begin{figure}
    \centering
    \includegraphics[width=1\linewidth]{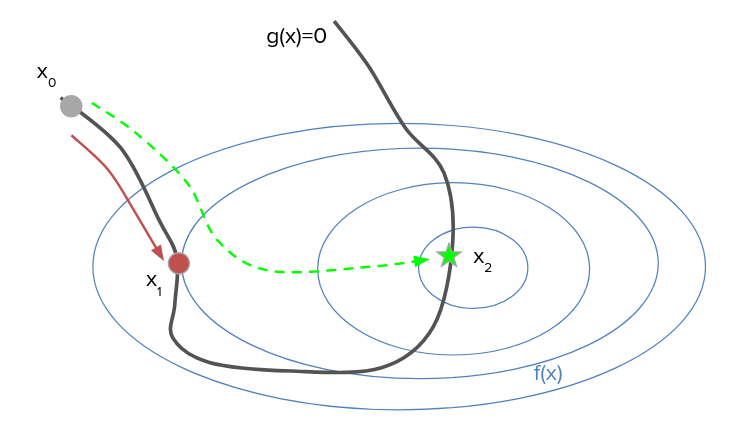}
    \caption{Sketch of optimization landscape (contours of $f$ in blue) with constraints (curve of $g(x)=0$ in black). Starting from $x_0$ and enforcing the constraints exactly at each step will follow the red path and end at $x_1$. If we allow ourselves to temporarily violate the constraints, we can follow the green path and arrive at the better solution $x_2$.}
    \label{fig:projection}
\end{figure}

This last point is illustrated in \autoref{fig:projection}. Points where the constraint curve $g(\mathbf{x})=0$ is tangent to the contours of $f$ are local minima of the constrained problem, where following the constraint curve will lead to higher objective function values, but moving in a direction to decrease the objective function will violate the constraints. If we enforce that the constraint is satisfied at each step, we will follow the path in red and stop at the local minimum $x_1$. However, if we are allowed to temporarily violate the constraints, we can make rapid progress towards the better local minimum $x_2$. 

Some may worry that not enforcing MHD equilibrium at each step may lead the optimizer in directions which are not dynamically accessible during an actual plasma discharge, where the plasma is assumed to be in equilibrium, however it is important to note that the path taken through phase space during optimization in general has no connection to the path taken during an actual discharge, and just because the optimizer finds a path that violates equilibrium does not mean there isn't a nearby path that satisfies it. Dynamical accessibility is an important aspect of stellarator optimization, but the expense of time dependent simulations preclude their use inside an optimization loop, so are generally limited to post-optimization validation and analysis.

\section{Methods for Constrained Optimization}
\label{sec:constrained}

We consider a general constrained optimization problem of the form:

\begin{equation}
\begin{aligned}
    \min_\mathbf{x} \quad & f(\mathbf{x}) \\
\textrm{s.t.} \quad &     \mathbf{g}_E(\mathbf{x}) = 0 \\
  & \mathbf{g}_I(\mathbf{x}) \ge 0    \\
  \label{eq:NRstdform}
\end{aligned}
\end{equation}

where as before $\mathbf{g}_E$ and $\mathbf{g}_I$ denote the equality and inequality constraints respectively. Most analysis and algorithms start from the Lagrangian for this problem which is

\begin{equation}
    \mathcal{L}(\mathbf{x}, \mathbf{y}_E, \mathbf{y}_I) = f(\mathbf{x}) - \mathbf{y}_E^T \mathbf{g}_E(\mathbf{x}) - \mathbf{y}_I^T \mathbf{g}_I(\mathbf{x})
\end{equation}

where $\mathbf{y}_E$ and $\mathbf{y}_I$ are the \textit{Lagrange multipliers} associated with the equality and inequality constraints respectively. The conditions for a given set $(\mathbf{x}^*, \mathbf{y}_E^*, \mathbf{y}_I^*)$ to be a solution to \autoref{eq:NRstdform} are the so called Karush-Kuhn-Tucker or KKT conditions \citep{nocedal2006numerical}:

\begin{subequations}
\label{eq:kkt}
\begin{align} 
    \nabla_\mathbf{x} \mathcal{L}(\mathbf{x}^*, \mathbf{y}_E^*, \mathbf{y}_I^*) = \nabla f - \mathbf{y}_E^T \nabla \mathbf{g}_E - \mathbf{y}_I^T \nabla \mathbf{g}_I &= 0 
    \label{eq:laggrad} \\
    \mathbf{g}_E(\mathbf{x}^*) &= 0 \\
    \mathbf{g}_I(\mathbf{x}^*) &\geq 0 \\
    \mathbf{y}_I^* &\geq 0 \\
    \mathbf{y}_E^* \cdot \mathbf{g}_E(\mathbf{x}^*) & = 0\\
    \mathbf{y}_I^* \cdot \mathbf{g}_I(\mathbf{x}^*) & = 0
    \label{eq:compslack}
\end{align}
\end{subequations}

The first condition is reminiscent of the first order optimality condition for an unconstrained problem, $\nabla f(\mathbf{x}^*) = 0$, with the objective function $f$ replaced by the Lagrangian $\mathcal{L}$, however it is important to note that we do not seek a minimum of the Lagrangian. It can be shown that solutions to \autoref{eq:kkt} always lie at a saddle point where the Lagrangian is minimized with respect to the primal variables $\mathbf{x}$ and maximized with respect to the dual variables $\mathbf{y}$.

It is useful to take a moment to consider the meaning and utility of the Lagrange multipliers $\mathbf{y}$. A first insight into the meaning of Lagrange multipliers can be found from the last of the KKT conditions in \autoref{eq:compslack}:

\begin{equation}
    \mathbf{y}_I^* \cdot \mathbf{g}_I(\mathbf{x}^*) = 0
\end{equation}

Meaning that either the constraint is zero, or it's associated multiplier is zero for each constraint. For a problem in the form of \autoref{eq:NRstdform}, an inequality constraint not equal to zero is said to be "non binding" or "not active", meaning that the constraint has no effect on the solution, and the minimum would not change if the constraint were removed. This is contrasted with a "binding" or "active" constraint that is preventing the solution from improving the objective $f$. This suggests that the Lagrange multipliers contain information about how much the constraints are affecting the solution. This can be made more explicit by considering the following problem:

\begin{equation}
\begin{aligned}
    \min_\mathbf{x} \quad & f(x) \\
\textrm{s.t.} \quad &     g(x) = 0 \\
  \label{eq:toy_problem}
\end{aligned}
\end{equation}

The Lagrangian for this problem is 

\begin{equation}
    \mathcal{L}(x,y) = f(x) - y g(x)
\end{equation}

We can now consider what happens if we perturb the constraint $g = 0$ to $g + \delta g = 0$. 
From \autoref{eq:laggrad} we see that at an optimum $x^*$ the change in the Lagrangian must be zero to first order, which gives

\begin{equation}
    \delta f = y \delta g
\end{equation}

Here we see that the Lagrange multipliers tell us how much our objective can improve by relaxing our constraints, or the so called "marginal cost" of enforcing the constraint $g = 0$. Additionally, the sign of the multipliers tells us which direction we should relax the constraints to improve the objective. 

We can use a similar approach to consider trade offs between conflicting constraints, where if we have constraints $g_1$ and $g_2$ that cannot both be satisfied, the ratio of their multipliers $y_1/y_2$ tells us how much we can expect to improve $g_1$ by relaxing $g_2$ and vice versa.

A number of methods have been developed for solving constrained optimization problems which we briefly summarize in the following subsections. More information can be found in standard texts on numerical optimization \citep{nocedal2006numerical, conn2000trustregion}

\subsection{Interior Point Methods}
A popular class of methods for dealing with inequality constraints are the so-called "interior point" methods, or sometimes referred to as "log barrier" methods. These methods solve problems of the form:

\begin{equation}
\begin{aligned}
    \min_\mathbf{x} \quad & f(\mathbf{x}) \\
\textrm{s.t.} \quad &     \mathbf{g}(\mathbf{x}) = 0 \\
   & 0 \le \mathbf{x} \\
\end{aligned}
\label{eq:ipopt}
\end{equation}

(any problem of the form of \autoref{eq:opt1} can be transformed into this form by appropriate change of variables). The bound on the variables is replaced by a logarithmic penalty term:

\begin{equation}
\begin{aligned}
    \min_\mathbf{x} \quad & f(\mathbf{x}) - \mu \sum_i \log{x_i} \\
\textrm{s.t.} \quad &     \mathbf{g}(\mathbf{x}) = 0 \\
\end{aligned}
\label{eq:ipopt1}
\end{equation}

This problem is then solved repeatedly with an equality constrained optimization method such as sequential quadratic programming (see the following section) for a decreasing sequence of barrier parameters $\mu$. For finite values of $\mu$, the log term pushes the solution into the interior of the feasible set (hence the name), and the solution is allowed to approach the boundary as the barrier gets sharper as $\mu$ approaches zero. 

The interior point method is the basis of a number of popular commercial and open source optimization codes \citep{biegler2009ipopt, wachter2006ipopt}, but we have found that in practice they do not perform well on the types of problems commonly encountered in stellarator optimization. We hypothesize this is due to two primary reasons:

\begin{enumerate}
    \item Stellarator optimization problems tend to be badly scaled due to spectral discretizations leading to different variables having exponentially more effect on the objective and constraints. In many cases this is not a problem, but the log barrier term can also lead to very poorly conditioned derivatives as the barrier is approached. Combined, this can significantly slow convergence or cause early termination due to loss of precision.
    \item Many interior point methods require any initial guess to be strictly feasible with respect to any inequality constraints. In many cases, an initial guess can be found from a near axis expansion or surrogate model, but it will generally not be a strictly feasible solution.
\end{enumerate}

\subsection{Sequential Quadratic Programming (SQP) Methods}

Another popular class of methods are based on sequentially minimizing an approximate model of the objective and constraints. Typically, the objective is modelled as quadratic and the constraints are modelled as linear, leading to a standard quadratic program. If the current estimate of the solution is $\mathbf{x}_k$, we seek the next iterate $\mathbf{x}_{k+1} = \mathbf{x}_k + \mathbf{p}$ by solving the following subproblem:

\begin{equation}
\begin{aligned}
    \min_\mathbf{p} \quad & \mathbf{p}^T H \mathbf{p} + \mathbf{g}^T \mathbf{p} \\
\textrm{s.t.} \quad &     A_E \mathbf{p} = -\mathbf{g}_E(\mathbf{x}_k) \\
 \mathbf{l} - \mathbf{g}_I(\mathbf{x}_k)  \le &A_I \mathbf{p} \le \mathbf{u} - \mathbf{g}_I(\mathbf{x}_k)
\end{aligned}
\label{eq:qp1}
\end{equation}

where $\mathbf{g}$ and $H$ are the gradient and Hessian of $f$ and $A_E$ and $A_I$ are the Jacobian matrices of the equality and inequality constraints $\mathbf{g}_E$ and $\mathbf{g}_I$.

One common issue with SQP methods is the possible inconsistency of the linearized constraints, such that the subproblem \autoref{eq:qp1} has no solution. This problem can be made worse by the inclusion of a trust region constraint, as is often used to ensure global convergence. To overcome this, the full step $\mathbf{p}$ is typically broken into two sub-steps, one of which attempts to achieve feasibility with respect to the linearized constraints as much as possible, while the other attempts to reduce the objective function. A merit function is then used to balance between improving the objective and decreasing the constraint violation \citep{conn2000trustregion}.

An alternative formulation, called "active set" methods, makes use of the fact that if one knows ahead of time which inequality constraints are active at the solution, then we can turn those into equality constraints and ignore all other inequality constraints, leading to a purely equality constrained problem which is often easier to solve. Active set methods make a guess for which constraints are active at the solution and iteratively update this set based on the results of sequential equality constrained optimizations. 

Similar to interior point methods, SQP methods form the core of a number of commercial and open source optimization codes \citep{gill2005snopt, johnson2021nlopt}. However, we similarly see that their performance is limited on the types of problems encountered in practice.

\begin{enumerate}
    \item If the Hessian matrix is not positive definite (which is common when far from the solution) subproblem \autoref{eq:qp1} is in general NP hard to solve (in that there is no polynomial time algorithm), so one must often be satisfied with an approximate solution which can significantly slow global convergence
    \item Active set methods can be very efficient for small problems, however the number of iterations required to determine the correct active set grows with the number of constraints, making active set methods inefficient for high dimensional problems.
\end{enumerate}

\subsection{Augmented Lagrangian Methods}
\label{sec:auglag}
For the quadratic penalty method considered in \autoref{eq:quad_penalty}, one can show that the constraint violation is inversely proportional to the penalty parameter:

\begin{equation}
    g_i(\mathbf{x}_k) \approx -y_i^* /\mu_k
\end{equation}

where $y_i^*$ is the true Lagrange multiplier (from the solution to \autoref{eq:kkt}) associated with the constraint $g_i$ and $\mu_k$ is the penalty parameter. We see that if we want to satisfy the constraints exactly, we must have $\mu \rightarrow \infty$, which as discussed can be numerically intractable. However, if we consider the "Augmented" Lagrangian

\begin{equation}
    \mathcal{L}_A(\mathbf{x}, \mathbf{y}, \mu) = f(\mathbf{x}) - \mathbf{y}^T \mathbf{g}(\mathbf{x}) + \frac{\mu}{2} ||\mathbf{g}(\mathbf{x})||^2 
\end{equation}

where $\mathbf{y}$ are estimates of the true Lagrange multipliers, one can show that now \citep{nocedal2006numerical}

\begin{equation}
    g_i(\mathbf{x}_k) \approx -(y_i^* - y_i^k) /\mu_k
\end{equation}

So that if our estimate of the true Lagrange multipliers is accurate, we can significantly reduce the constraint violation without requiring the penalty parameter to grow without bound. This also gives us a rule to update our estimate of the multipliers:

\begin{equation}
    y_i^{k+1} =  y_i^k - \mu_k g_i(\mathbf{x}_k)
\end{equation}

Minimizing the augmented Lagrangian with fixed $\mathbf{y}$ and $\mu$ is an unconstrained or bound constrained problem, for which standard gradient or Newton based methods can be used. From this solution we obtain an approximation to the true solution of the constrained problem. We can then update the multipliers and penalty term and re-optimize, eventually converging to the true solution. This is the basic method used in several widely used codes \citep{conn2013lancelot, fowkes2023galahad} and can also be used as a generic technique to handle constraints for algorithms that otherwise would only be able to solve unconstrained problems \citep{johnson2021nlopt}.

One can also construct an augmented Lagrangian for a least squares problem. Because $L(\mathbf{x}, \mathbf{y}, \mu)$ is only ever minimized with respect to $\mathbf{x}$, one can add any quantity independent of $\mathbf{x}$ without affecting the result. In particular, if we add $||\mathbf{y}||^2/2\mu$, we get

\begin{equation}
    \mathcal{L}_A(\mathbf{x}, \mathbf{y}, \mu) = \frac{1}{2} ||\mathbf{f}(\mathbf{x})||^2 - \mathbf{y}^T \mathbf{g}(\mathbf{x}) + \mu/2 ||\mathbf{g}(\mathbf{x})||^2 + ||\mathbf{y}||^2/2\mu
\end{equation}

We can then complete the square to obtain

\begin{equation}
    \mathcal{L}_A(\mathbf{x}, \mathbf{y}, \mu) = \frac{1}{2} ||\mathbf{f}(\mathbf{x})||^2 + \frac{1}{2} ||-\mathbf{y}/\sqrt{\mu} + \sqrt{\mu}\mathbf{g}(\mathbf{x})||^2
    \label{eq:lsq_auglag}
\end{equation}

This allows us to use standard unconstrained or bound constrained least squares routines to solve the augmented Lagrangian subproblem, which has a number of benefits:

\begin{enumerate}
    \item Super-linear convergence can be obtained using only first derivatives of $\mathbf{f}$ and $\mathbf{g}$ (as opposed to standard Newton methods which require second derivatives of the objective and constraints).
    \item The quadratic subproblem that arises in least squares is always convex, for which the exact solution can be found efficiently.
\end{enumerate}

\section{Constrained Optimization in DESC}
\label{sec:indesc}

As previously mentioned there are a wide variety of commercial and open source codes for constrained optimization, many of which have been interfaced to the DESC stellarator optimization code \citep{dudt2020desc, panici2023desc, conlin2023desc, dudt2023desc, desc_git}. In practice, we have found that most of the existing methods perform somewhat poorly on the types of problems encountered in stellarator optimization. We hypothesize this is for a number of reasons: 
\begin{enumerate}
    \item Many commercial solvers assume a particular structure for the problem, such as quadratic, conic, and semi-definite programming solvers. Others rely on convexity, partial separability, or other conditions that generally don't hold in stellarator optimization. 
    \item Many codes are designed and optimized for solving very high dimensional problems ($O(10^6)$ variables or more) but with a high degree of sparsity, and make extensive use of sparse and iterative linear algebra methods. In stellarator optimization problems, we generally do not have significant sparsity due to the wide use of spectral methods, making these methods inefficient.
    \item The types of problems in stellarator optimization are often poorly scaled, due to a wide range of scales in physical units, and due to spectral discretizations. This can lead to failure of iterative linear algebra subproblems without a proper preconditioner (which may not be easy to determine) and can also slow progress of the overall optimization because the optimization landscape is strongly anisotropic (meaning the objective is much more sensitive to changes in certain directions in parameter space than others, such as a long narrow valley).
    \item No existing methods that we are aware of are able to take advantage of GPU acceleration for the underlying linear algebra, limiting the overall efficiency on modern heterogeneous computing environments.
\end{enumerate}

Because of this, we have implemented a new optimizer based on the least squares augmented Lagrangian formalism from \autoref{sec:auglag}, and a sketch of the algorithm is found in Algorithm \ref{alg:auglag}. 

\begin{algorithm}
\caption{Augmented Lagrangian Algorithm (adapted from \citep{nocedal2006numerical, conn2000trustregion} }
\label{alg:auglag}
\begin{algorithmic}
\Require $\mu_0 > 0$, $\omega_* > 0$, $\eta_* > 0$
\State $\mu_k \gets \mu_0$
\State $\omega_k \gets 1/\mu_0$
\State $\eta_k \gets 1/\mu_0^{0.1}$
\State $k \gets 1$
\While{$||\nabla \mathcal{L}_A(\mathbf{x}_k, \mathbf{y}_k, \mu_k)|| > \omega_*$ or $||\mathbf{g}(\mathbf{x}_k, \mathbf{y}_k, \mu_k)|| > \eta_*$}
\State $\mathbf{x}_{k+1} \gets \argmin_\mathbf{x} \mathcal{L}_A(\mathbf{x}_k, \mathbf{y}_k, \mu_k)$ \Comment{using unconstrained or bound constrained method, solved to tolerance $\omega_k$}
\If{$||\mathbf{g}(\mathbf{x}_k, \mathbf{y}_k, \mu_k)|| < \eta_k$}
\Comment{Constraints are making good progress, update multipliers}
    \State $\mathbf{y}_{k+1} \gets \mathbf{y}_k - \mu_k \mathbf{g}(\mathbf{x}_{k+1})$
    \State $\mu_{k+1} \gets \mu_k$
    \State $\omega_{k+1} \gets \omega_k / \mu_{k+1}$
    \State $\eta{k+1} \gets \eta_k / \mu_{k+1}$
    
\Else
\Comment{Need to improve constraint satisfaction, increase penalty term}
    \State $\mathbf{y}_{k+1} \gets \mathbf{y}_k$
    \State $\mu_{k+1} \gets 10 \mu_k$
    \State $\omega_{k+1} \gets 1 / \mu_{k+1}$
    \State $\eta{k+1} \gets 1 / \mu_{k+1}$
\EndIf
\State $k \gets k+1$
\EndWhile
\end{algorithmic}
\end{algorithm}

The precise rules for updating the tolerances and penalty parameters can be varied, see \citep{conn2000trustregion} for a more complex version, though we have found that in practice the results and performance do not significantly depend on the choice of hyper-parameters. These rules are effectively automatically finding the correct weights to manage the trade-off between improving the objective value and constraint feasibility that would normally require extensive user trial and error.

\section{Examples}
\label{sec:examples}
\begin{figure}
    \centering
    \includegraphics[width=0.5\linewidth]{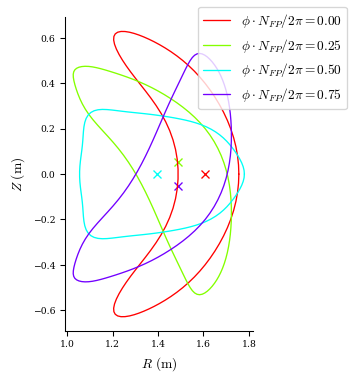}
    \caption{Plasma boundary of NCSX at several cross sections, with the "bean" section at $\phi=0$.}
    \label{fig:ncsx}
\end{figure}

Many optimized stellarators have very strongly shaped boundaries, and a "bean" shaped cross section is quite common, appearing in designs such as NCSX (shown in \autoref{fig:ncsx}), W7-X, ARIES-CS, and many recent designs such as the "precise" quasisymmetric equilibria of Landreman and Paul \citep{landreman2022precise}. These strongly indented shapes can be difficult to create with external coils or magnets, often requiring coils that are very close on the inboard concave side \citep{kappel2024magnetic, landreman2016efficient}. 

We can attempt to alleviate this problem with constrained optimization by properly constraining the curvature of the surface. We consider a surface parameterized by poloidal angle $\theta$ and toroidal angle $\zeta$, with parameterization $\mathbf{r}(\theta, \zeta)$. The covariant basis vectors are given by $\mathbf{r}_\alpha (\theta, \zeta) = \partial_\alpha \mathbf{r}(\theta, \zeta)$ for $\alpha \in (\theta, \zeta)$. From these we can construct the "first fundamental form" \citep{docarmo2016differential}

\begin{equation}
    \mathcal{F} = \begin{pmatrix}
        E & F \\
        F & G \\
    \end{pmatrix}
\end{equation}

where the coefficients are dot products of the first derivative of the position vector:

\begin{align}
    E &= \mathbf{r}_\theta \cdot \mathbf{r}_\theta \\
    F &= \mathbf{r}_\theta \cdot \mathbf{r}_\zeta \\
    G &= \mathbf{r}_\zeta \cdot \mathbf{r}_\zeta \\
\end{align}

Similarly we can define the "second fundamental form"

\begin{equation}
    \mathcal{S} = \begin{pmatrix}
        L & M \\
        M & N \\
    \end{pmatrix}
\end{equation}

Where the coefficients are given by the projection of the second derivatives of the position vector onto the surface normal:

\begin{align}
    L &= \mathbf{r}_{\theta \theta} \cdot \mathbf{n} \\
    M &= \mathbf{r}_{\theta \zeta} \cdot \mathbf{n} \\
    N &= \mathbf{r}_{\zeta \zeta} \cdot \mathbf{n} \\
\end{align}

We can then compute the principal curvatures $\kappa_1$ and $\kappa_2$ of the surface as the solutions to the generalized eigenvalue problem 

\begin{equation}
    \mathcal{S}\mathbf{u} = \kappa \mathcal{F} \mathbf{u}
\end{equation}

where the eigenvectors $\mathbf{u}$ are called the principal directions. The principal curvatures are the maximum and minimum of the inverse radii of curvature of all the curves that are tangent to the surface at a given point, with the sign convention that positive curvature means the surface curves towards the normal vector.

On the inboard side of a toroidal surface, we expect at least one of the principal curvatures to be positive because the surface curves towards the normal in the toroidal direction, while at least one should be negative on the outboard side. Indenting of the poloidal cross section characteristic of the bean shape would manifest as both principal curvatures being positive on the inboard side, and both negative on the outboard side.

To prevent strong indentation, we will use the mean curvature, defined as $H = 1/2(\kappa_1 + \kappa_2)$ and enforce that the mean curvature is everywhere negative. On the outboard side, where both curvatures are already negative this should have minimal effect, while on the inboard it forces any indent to have a larger radius of curvature than the local major radius.

In addition to this constraint on the curvature of the plasma surface, we also include constraints to enforce MHD equilibrium, fix the major radius $R_0$, aspect ratio $R_0/a$, total enclosed volume $V$ (which we fix equal to the volume of the initial guess, a circular torus) and include inequality constraints on the rotational transform profile $\iota$. We also fix the profiles of pressure $p$ and current $j$, and the total enclosed flux $\Psi$ to give an average field strength of 1T. In summary, our constraints are

\begin{align*}
    R_0 &= 1~\mathrm{m} \\
    R_0/a &= 6 \\
    V &= V_{init} \sim 0.55~\mathrm{m}^3 \\
    0.43 < \iota(\rho) &< 0.5 \\
    H(\theta, \zeta) &< 0~\mathrm{m}^{-1}\\
    p(\rho) &= 0~\mathrm{Pa} \\
    j(\rho) &= 0~\mathrm{A} \\
    \Psi &= 0.087~\mathrm{Tm}^2 \\
    \mathbf{J} \times \mathbf{B} - \nabla p &= 0 \\
\end{align*}

As our objective, we use the "two-term" metric $f_C$ for quasisymmetry (section 2.3.2 of \citep{dudt2023desc}), targeting quasi-axisymmetry at 10 surfaces evenly spaced throughout the volume.

\begin{equation}
    f_C = (M \iota - N) (\mathbf{B} \times \nabla \psi) \cdot \nabla B - (M G + N I) \mathbf{B} \cdot \nabla B
\end{equation}

Here, $\mathbf{B}$ is the magnetic field with magnitude $B$, $M$ and $N$ give the type of quasisymmetry (for QA, $M=1$, $N=0$), and $G$ and $I$ are the poloidal and toroidal covariant components of the field in Boozer coordinates.

\begin{figure}
    \centering
    \includegraphics[width=0.5\linewidth]{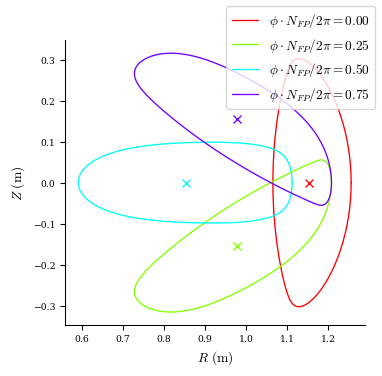}
    \caption{Plasma boundary of new optimized configuration with constraint on surface curvature.}
    \label{fig:opt1}
\end{figure}

\begin{figure}
    \centering
    \includegraphics[width=0.5\linewidth]{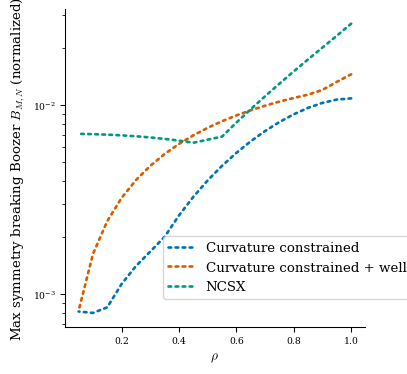}
    \caption{Quasisymmetry error measured by the maximum amplitude of the symmetry breaking Boozer harmonics for the newly optimized configuration with and without magnetic well constraint, compared to NCSX.}
    \label{fig:qs1}
\end{figure}

The plasma boundary after optimization is shown in \autoref{fig:opt1}, and we can see that we have successfully avoided any concave regions that may be difficult to produce with external coils. In \autoref{fig:qs1} we also see that the level of quasisymmetry in this new configuration compares quite well with similar optimized designs such as NCSX. However, upon further analysis we find that this new configuration has a negative magnetic well, indicating it is MHD unstable.

We can attempt to rectify this by adding a constraint that the magnetic well be positive and re-running the optimization. Doing this, we find that the optimizer does not converge, indicating that the constraints cannot all be satisfied at the same time. This is perhaps expected, as empirical evidence suggests that bean-like cross sections can be favorable for MHD stability \citep{nuhrenberg1986stable, nuhrenberg2010development}, though may not be strictly required from a near axis formalism \citep{rodriguez2023magnetohydrodynamic}.

\begin{figure}
    \centering
    \includegraphics[width=1\linewidth]{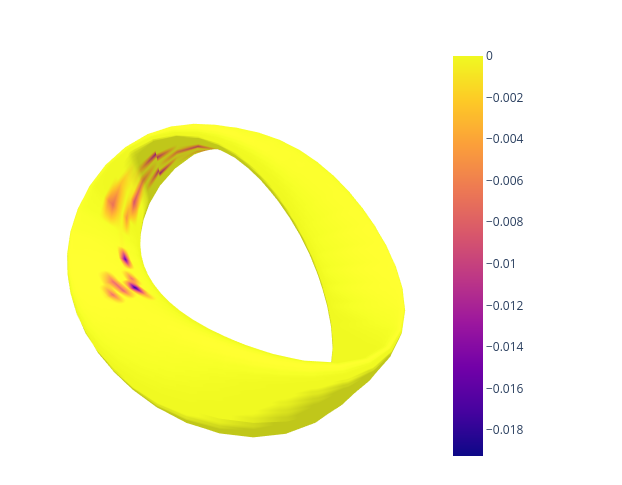}
    \caption{Lagrange multipliers for the curvature constraint plotted on the surface. The dark areas indicate places where the constraint is binding, and the optimizer wants to indent the boundary in those locations to improve the magnetic well.}
    \label{fig:multipliers}
\end{figure}

However, in this case, we can use the estimated Lagrange multipliers to get an idea of which constraints are the most limiting. Because the curvature constraint is enforced point wise in real space, the Lagrange multipliers also have a local characteristic, and we can plot them over the surface to see where the optimizer wanted to indent the surface, as shown in \autoref{fig:multipliers}. As expected, the curvature constraint is most prominent on the inboard side, and the negative sign of the multipliers indicates that increasing the curvature in those regions will be favorable for stability.

\begin{figure}
    \centering
    \includegraphics[width=0.5\linewidth]{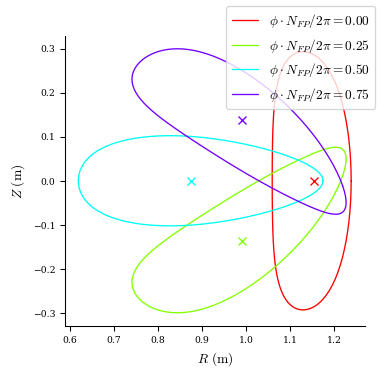}
    \caption{Plasma boundary of new optimized configuration with relaxed curvature constraint and stable magnetic well.}
    \label{fig:opt2}
\end{figure}

Using this information, we re-run the optimization with the magnetic well constraint, but reduce the upper bound on mean curvature from 0 m$^{-1}$ to 0.5 m$^{-1}$. With this relaxed constraint we are able to achieve a stable magnetic well, while maintaining most of the benefits of the curvature constraint, as shown in \autoref{fig:opt2}. This is not without some cost, as we see in \autoref{fig:qs1} that the level of quasisymmetry for the stable configuration is degraded somewhat compared to the unstable one, though still at reasonable levels compared to other optimized configurations such as NCSX.

\section{Conclusion}

In this work we have described the many ways that constraints naturally arise in stellarator optimization problems and surveyed a number of methods to solve constrained optimization problems, including methods that have not been widely applied in the field thus far. We have further developed and implemented a new constrained optimization algorithm in the DESC code and demonstrated its usefulness in designing a newly optimized equilibrium that achieves good quasisymmetry and a stable magnetic well without strong shaping of the plasma boundary. We are hopeful that these new techniques will open new avenues of possibility for the design and optimization of future stellarators, and significantly reduce the amount of guesswork and "black magic" required when designing objectives for optimization, as well as offering insights into trade offs between different objectives and constraints.

\section*{Funding}
This work was supported by the U.S. Department of Energy under contract numbers DE-AC02-09CH11466, DE- SC0022005 and Field Work Proposal No. 1019. The United States Government retains a non-exclusive, paid-up, irrevocable, world-wide license to publish or reproduce the published form of this manuscript, or allow others to do so, for United States Government purposes.

\section*{Declaration of interests}
The authors report no conflict of interest.

\section*{Data availability statement}
The source code to generate the results and plots in this study are openly available in DESC at https://github.com/PlasmaControl/DESC or http://doi.org/10.5281/zenodo.4876504

\section*{Author ORCID}
R. Conlin, https://orcid.org/0000-0001-8366-2111; D. Dudt, https://orcid.org/0000-0002-4557-3529; D. Panici, https://orcid.org/0000-0003-0736-4360; E. Kolemen, https://orcid.org/0000-0003-4212-3247

\bibliographystyle{jpp}

\bibliography{bibliography}

\end{document}